# Generalized Blockmodeling of Valued Networks[♦]

Aleš Žiberna[*]

*University of Ljubljana*

## Abstract

The paper presents several approaches to generalized blockmodeling of valued networks, where values of the ties are assumed to be measured on at least interval scale. The first approach is a straightforward generalization of the generalized blockmodeling of binary networks (Doreian et al., 2005) to valued blockmodeling. The second approach is homogeneity blockmodeling. The basic idea of homogeneity blockmodeling is that the inconsistency of an empirical block with its ideal block can be measured by within block variability of appropriate values. New ideal blocks appropriate for blockmodeling of valued networks are presented together with definitions of their block inconsistencies.

## 1  Introduction

The paper presents several approaches to generalized blockmodeling of valued networks. All these approaches are implemented in an R package `blockmodeling` (Žiberna, 2006). Unless explicitly stated, the term "valued networks" is used for valued networks where values of the ties are assumed to be measured on at least interval scale. Valued networks not measured on at least interval scale (for example categorical) are not covered by this paper.

"Blockmodeling tools were developed to partition network actors (units) into clusters, called positions, and, at the same time, to partition the set of ties into blocks that are defined by the positions (see Lorrain and White (1971), Breiger et al. (1975), Burt (1976) for the foundational statements)" (Doreian et al., 2004). It could be said that blockmodeling seeks clusters of equivalent units based on a selected definition of equivalence.

Doreian et al. (2005) state that there are three main characteristics of generalized blockmodeling (in comparison to what they call conventional blockmodeling):
- o  A direct approach is taken to optimization (the algorithm works directly with network data and does not transform them into some other form);
- o  A much broader set of ideal blocks is used instead of a few equivalence types;
- o  The model can be pre-specified (not only the allowed ideal blocks, but also their locations within a blockmodel).

In this paper several types of generalized blockmodeling are distinguished: binary (relational) blockmodeling [1], valued blockmodeling and homogeneity blockmodeling. In the case of

---

[♦] The author would like to thank Anuška Ferligoj, Patrick Doreian, and Vladimir Batagelj whose suggestions led to improvements in the paper.
[*] Aleš Žiberna, Faculty of Social Sciences, University of Ljubljana, Kardeljeva ploščad 5, 1000 Ljubljana, Slovenia; E-mail: ales.ziberna@gmail.com; tel: +38631345468.
[1] The term binary blockmodeling type is used for generalized blockmodeling of binary networks, which was thoroughly presented by Doreian et al. (2005).

homogeneity blockmodeling, two types are defined: sum of squares blockmodeling and absolute deviations blockmodeling. The aim of this paper is to discuss blockmodeling of valued networks. Therefore, binary blockmodeling is considered only as a basis to develop appropriate approaches to blockmodeling of valued networks. It is also used for comparison with proposed approaches.

The most important difference between the three main types of generalized blockmodeling is an appropriate definition of the criterion function, which measures the inconsistencies of the empirical blocks with the ideal ones. The criterion functions have to consider the fact that values of the ties in a valued network are measured on at least interval scale. Therefore, ideal blocks must also be redefined. This is discussed in Sections 4 and 5. Ideal blocks of different types of generalized blockmodeling can not be used together in the same blockmodel.

Binary blockmodeling analyzes only binary[2] networks and its criterion function measures block inconsistencies in principle with the number of errors. The other two main types of generalized blockmodeling analyze valued networks. The criterion function of the valued blockmodeling measures block inconsistencies as the deviation of appropriate values from either 0 or the value determined by the parameter $m$. The criterion functions of both types of homogeneity blockmodeling measure the block inconsistencies with the variability of appropriate values.

Other blockmodeling approaches can be also used for valued networks. For structural equivalence these approaches include both indirect and direct approaches (e. g. see Batagelj et al., 1992b; Breiger et al. 1975). Two direct approaches are appropriate: an approach suggested by Breiger and Mohr (2004) based on log-linear models and an approach suggested by Borgatti and Everett (1992b), which is similar to homogeneity blockmodeling. Doreian and Mrvar (1996, 160-162) also presented an approach for the analysis of valued signed graphs[3]. The part of their criterion function that corresponds to the "positive error" is the same as inconsistency for null blocks in valued blockmodeling.

For regular equivalence, only two[4] versions of the REGE (White, 1985a; 1985b) algorithm exist for blockmodeling of valued networks. The comparison between the approaches presented in this paper and REGE is discussed in Žiberna (2005). A number of other approaches exist. However since they search for different types of equivalences than those implemented in the approaches suggested in this paper, they are not mentioned here.

## 2 Equivalences

As stated previously, blockmodeling seeks clusters of equivalent units based on some notion of equivalence. There are several types of well known equivalences in social network analysis, all originally defined for binary networks. The first one is structural equivalence, where units are structurally equivalent if they have identical ties to the rest of the network (and themselves) (Lorrain and White, 1971). The other well known type of equivalence is regular equivalence. Regular equivalence is an attempt to generalize structural equivalence and as such includes structural equivalence as a special case. The units are regularly equivalent if they are connected in the same way to equivalent others (White and Reitz, 1983). Doreian et al. (1994) introduced the concept of generalized equivalence, which is defined by the set of allowed ideal blocks.

---

[2] Batagelj (1997: 148) presented a set of averaging rules for assigning values to the ties in the blockmodel obtained by (binary) generalized blockmodeling based on types of blocks and values of the ties. Therefore, this approach also takes into account values of ties. However it does so only for assigning values to the ties in the blockmodel. The partition and the blockmodel are based only on the binary (relational) data.
[3] Signed graphs are not discussed in this paper.
[4] There exists a third version of REGE, the CATREGE algorithm (Borgatti and Everett, 1993), which is however designed for categorically valued networks. This type of valued networks is not covered by this paper.

Let us first define some notations:

- The network N = (*U*, *R*), where *U* is a set of all units *U* = ($u_1$, $u_2$, ..., $u_n$) and *R* is the relation between these units $R \subseteq U \times U$.
- In generalized blockmodeling, a relation *R* is usually represented by a valued matrix R with elements [$r_{ij}$], where $r_{ij}$ indicates the value (or weight) of the arc from unit *i* to unit *j*; $r: R \to \mathbb{R}$, $r_{ij} = \begin{cases} r(i,j), (i,j) \in R \\ 0, \text{otherwise} \end{cases}$
- $C_i$ is a cluster of units.
- **C** = {$C_1, C_2, ..., C_n$} is a partition of the set *U*; $\bigcup_{i=1}^{n} C_i = U$ ; $C_i \cap C_j = 0, i \neq j$.
- $\Phi$ is a set of all feasible partitions.
- A partition **C** also partitions the relation *R* into blocks; $R(C_i, C_j) = R \cap C_i \times C_j$. Each such block consists of units belonging to clusters $C_i$ and $C_j$ and all arcs leading from cluster $C_i$ to cluster $C_j$. If *i* = *j*, a block $R(C_i, C_i)$ is called a diagonal block.
- Let $T(C_i, C_j)$ denote a set of all ideal blocks, corresponding to an empirical block $R(C_i, C_j)$. Ideal blocks are defined in Table 1 in Section 4.
- *f* is a function that assigns to a valued vector of length n a real value; $f: \mathbb{R}^n \to \mathbb{R}$. For example, this function can be *mean, maximum, sum, ...*

Borgatti and Everett (1992b – Definition 5) gave a formal definition of structural equivalence for valued networks. However, this definition of structural equivalence has exactly the shortcomings that the authors (Borgatti and Everett, 1992a) criticize in the original definition of structural equivalence given by Lorrain and White (1971). These shortcomings are shown in the fact that by this definition in a network without loops, two units that are connected can not be structurally equivalent. A definition without these shortcomings was given by Batagelj et al. (1992b) and this one is used in this paper. Although the definition was formulated for binary networks, it can also be used for valued networks, as already noticed noted by Batagelj et. al (2004).

Definition: suppose that $\equiv$ is an equivalence relation on *U* then $\equiv$ is a *structural equivalence* if and only if for all *a*, *b* $\in$ *U*, *a* $\equiv$ *b* implies:
1. $r_{bi} = r_{ai}$ for all *i* $\in$ *U* \ {*a*, *b*},
2. $r_{ib} = r_{ia}$ for all *i* $\in$ *U* \ {*a*, *b*},
3. $r_{bb} = r_{aa}$, and
4. $r_{ab} = r_{ba}$.

Borgatti and Everett (1992b – Definition 6) also provided a formal definition of regular equivalence for valued networks. Alternative definitions of regular equivalence for valued networks measured on at least interval scale can be formulated from:
- Two algorithms, REGGE and REGDI (White, 1985a; 1985b), for measuring the similarities and dissimilarities of units in terms of regular equivalence.
- Ideas for defining block inconsistencies (generalized blockmodeling approach) for valued networks presented by Batagelj and Ferligoj (2000) that can be also used for regular equivalence.

However, none of these definitions are used in this paper. For valued networks, another type of equivalence could be useful. For now, let us call it *f*-regular (or *function*-regular) equivalence. This equivalence is used in place of regular equivalence in the paper. This definition can be formally best presented in matrix terms and is given below. The definition is given for single relation networks; however, it can be generalized to multi-relational networks by demanding that the definition holds for all relations.

**Definition:** Suppose $\equiv$ is an equivalence relation on $U$ that induces (or corresponds to) partition **C**. Then $\equiv$ is an *f-regular equivalence* (where *f* is a selected function, like *sum*, *maximum*, *mean* …) if and only if for all $a, b \in U$ and all $X \in \mathbf{C}$, $a \equiv b$ implies:

1. $\sum_{i \in X} f(r_{ai}) = \sum_{i \in X} f(r_{bi})$ and
2. $\sum_{i \in X} f(r_{ia}) = \sum_{i \in X} f(r_{ib})$.

If the function *f* is *maximum*, we get the definition of the regular equivalence for valued networks that can be defined based on ideas of Batagelj and Ferligoj (2000). However, other functions are also appropriate, especially *sum* and *mean*. *Sum*-regular equivalence would, for binary networks, correspond to exact coloration as presented by Everett and Borgatti (1994). The comparison with the definitions that can be formulated based on two algorithms developed by White (1985a; 1985b) can be found in Žiberna (2005).

This could be considered as another alternative definition of the regular equivalence for valued networks, or probably better, as a new type of equivalence. The definition is not meant to be a strict generalization of the regular equivalence to valued networks. However, it tries to capture the idea that it is not necessary for equivalent units to be equivalently connected to an individual unit, but only to a group of equivalent units.

In order to find partitions that match (or almost match) a selected equivalence, the equivalence must be operationalized (by an equivalence detector). Batagelj et al. (1992a,b) introduced the criterion (fit) function that measures how the data fit an equivalence as a means of operationalizing equivalences in direct approaches. In the context of generalized blockmodeling, operationalization of a specific equivalence is done by specifying ideal blocks and measures of block inconsistencies of empirical blocks with these ideal blocks. For binary networks, this was already done by Batagelj et al. (1992b) for structural equivalence, by Batagelj et al. (1992a) for regular equivalence, and by Doreian et al. (1994) for generalized equivalence. Batagelj and Ferligoj (2000) presented some ideas for operationalizing these equivalences for valued networks. Ideal blocks and measures of block inconsistencies of empirical blocks with these ideal blocks are presented in Section 4 for valued blockmodeling and in Section 5 for homogeneity blockmodeling. Borgatti and Everett (1992b) also operationalized structural equivalence for valued networks.

## 3   Criterion function

A common component of all types of generalized blockmodeling is a basic criterion function. The only part of the criterion function that changes among different types of generalized blockmodeling is the part where inconsistencies with ideal blocks are computed. The rest of the criterion function is the same as in generalized blockmodeling of binary networks (binary blockmodeling) as presented by Doreian et al. (2005). Some properties of part of the criterion function common to all types of generalized blockmodeling are discussed in this section.

$\delta(R(C_i,C_j),T)$ measures the deviation (the inconsistency) of the empirical block $R(C_i, C_j)$ from (with) the ideal block $T \in T(C_i, C_j)$. This term can also be normalized to exclude the effect of block size by dividing it by the number of cells in the block.

Block inconsistency $p(C_i, C_j)$ is defined as $p(C_i,C_j) = \min_{B \in B(C_i,C_j)} \delta(R(C_i,C_j),T)$.

The total inconsistency $P(\mathbf{C})$ of a partition $\mathbf{C}$ can be expressed as sum of inconsistencies within each block (block inconsistencies) across all blocks: $P(\mathbf{C}) = \sum_{C_i,C_j \in \mathbf{C}} p(C_i,C_j)$.

A criterion function is compatible with a definition of equivalence if $P(\mathbf{C}) = 0$ if and only if $\mathbf{C}$ induces that equivalence.

These definitions hold for all types of generalized blockmodeling of valued networks. The difference between different types of generalized blockmodeling is in the descriptions of ideal bocks $T(C_i, C_j)$ and in the definitions of block inconsistencies ($\delta(R(C_i,C_j),T)$). For valued and homogeneity blockmodeling, these had to be adapted to valued networks. This is covered in Sections 4 and 5.

## 4 Generalization of binary blockmodeling to valued blockmodeling

The approach presented in this section was inspired by the fact that in the past, when valued networks were analyzed using generalized blockmodeling (binary blockmodeling), they were first converted to binary networks. This was done by recoding values over (or equal to) a certain threshold (often 1) into ones and the other into zeros (examples can be found in Doreian et al., 2005). Analyzing a network in such a way causes the loss of a considerable amount of information. The approach presented in this section (valued blockmodeling) reduces the amount of information lost, although some loss usually still occurs. Information about the values of ties (or sometimes values of function *f* over certain values) above *m* (a parameter that is defined later) is lost. If *m* is set to a sufficiently high[5] value, no information is lost. However, this might not be appropriate in many networks, since it might cause almost all or even all blocks to be declared null. It should be also noted that both approaches mentioned above[6] do not search for the structural or *f*-regular equivalence for valued networks, defined in Section 2. The equivalences they search for are defined using ideal blocks (presented in the following subsection). In the case of binary blockmodeling, the threshold used in the transformation of a valued network into a binary network must also be considered.

In the remainder of this section, the ideal blocks for valued blockmodeling and the inconsistencies of empirical blocks with these ideal blocks are defined. This is done by generalization of ideal blocks for binary blockmodeling to valued networks. At the end of this section, the parameter *m*, introduced in the generalization to valued networks, is discussed.

**Ideal blocks**

We first look at descriptions of ideal blocks for binary blockmodeling (for binary networks). They are presented in Table 1 (for now, look at only the first two columns).

---

[5] What is "sufficiently high value" depends on the considered ideal blocks. In case of null blocks, no information is lost. For complete, row- and column-dominant and row and column-functional blocks, no information is lost if *m* is equal to or larger than the maximum of values in considered empirical blocks. For *f*-regular, row- and columns-*f*-regular blocks, no information is lost, if no value of function *f* over all rows (row-*f*-regular), columns (column-*f*-regular) or both (*f*-regular) is larger than *m*.

[6] Binary blockmodeling when used on valued networks and valued blockmodeling, if loss of information occurs.

All ideal blocks can be described using only three types of conditions:
1. **a certain cell must be (at least) 1,**
2. **a certain cell must be 0 and**
3. at least 1 cell in each row (or column) must be 1 or, to put it differently, **the $f$ over each row (or column) must be at least 1,** where $f$ is some function over $a$ that has the property $f(a) \geq \max(a)$, and $a$ is a valued vector.

*Table 1: Characterizations of ideal blocks*

| Ideal block with "label" | Description for binary blockmodeling[7] | Description for valued blockmodeling | Description for homogeneity blockmodeling |
|---|---|---|---|
| **null "null"** | all 0 ♠ | all 0 ♣ | all 0 ♥ |
| **complete "com"** | all 1 * | all values at least $m$ * | all equal ♦ |
| **row-dominant "rdo"** | there exists an all 1 row * | there exists a row where all values are at least $m$ * | there exists a row where all values are equal ♥● |
| **col-dominant "cdo"** | there exists an all 1 column * | there exists a column where all values are at least $m$ * | there exists a column where all values are equal ♥● |
| **row(-$f$)-regular "rre"** | there exists at least one 1 in each row | the $f$ over each row is at least $m$ | $f$ over all rows equal |
| **column(-$f$)-regular "cre"** | there exists at least one 1 in each column | the $f$ over each column is at least $m$ | $f$ over all columns equal |
| **($f$-)regular "reg"** | there exists at least one 1 in each row and each column | the $f$ over each row and each column is at least $m$ | $f$ over all rows and all columns separately equal |
| **row-functional "rfn"** | there exists exactly one 1 in each row | there exists exactly one tie with value at least $m$ in each row, all other 0 | *max* over all rows equal, all other values 0 ● |
| **column-functional "cfn"** | there exists exactly one 1 in each column | there exists exactly one tie with value at least $m$ in each column, all other 0 | *max* over all rows equal, all other values 0 ● |

Legend:
* - an exception may be cells on the diagonal, where then all cells should have value 0
♠ - an exception may be cells on the diagonal, where then all cells should have value 1
♣ - an exception may be cells on the diagonal, where then all cells should have value at least $m$
♥ - an exception may be cells on the diagonal, where then all values should be equal
♦ - cells on the diagonal may be treated separately – their values should all be equal, however can be different from the values of the off-diagonal cells
● - These descriptions/definitions of ideal blocks may change in the future. There are presented here only as suggestions and are not yet evaluated.

At least one of these three conditions can be found in each ideal block (of binary blockmodeling). The three of them together are enough to describe (with the correct specification of "certain cell") all ideal blocks presented in Table 1 and compute their inconsistencies. The block inconsistency is the weighted sum of the number of times each condition is broken (which applies to a certain block)[8]. The term weighted sum is used, since sometimes the block inconsistency or part of it is multiplied by the number of cells in a given row or column. If we could generalize these three conditions to valued networks, we would generalize the ideal blocks to valued networks.

---
[7] Doreian et al., 2005: 223.

We only need to replace the ones (1) in the conditions with an appropriate parameter (denoted by *m*) to generalize these three conditions to valued networks. The new conditions can then be written as:
1. **a certain cell must be (at least) *m*,**
2. **a certain cell must be 0, and**
3. **the *f* over each row (or column) must be at least *m*,** where *f* is again some function over *a* that has the property $f(a) \geq \max(a)$, and *a* is a valued vector.

Based on the generalization of these conditions and descriptions of ideal blocks for binary networks (presented in the second column of Table 1) the description of ideal blocks for valued networks presented in the third column of Table 1 can be derived. The fourth column is discussed later in Section 5. It should be noted that the ideal blocks for valued blockmodeling do not perfectly match structural and *f*-regular equivalence as they were defined in Section 2. As a result, criterion function for valued blockmodeling are not fully compatible with these definitions of structural and *f*-regular equivalence. The criterion function is compatible with the equivalence defined by a selection of allowed ideal blocks defined in the third column of Table 1.

On the basis of these descriptions, deviations (inconsistencies) of empirical blocks from (with) ideal blocks can be defined. In Table 2 these definitions are presented.

One nice property of this generalization is that if we have a binary network and set *m* to 1, these inconsistencies match those of binary blockmodeling. Therefore, binary blockmodeling can be seen as a special case of valued blockmodeling. Actually, even functions that do not comply with the property $f(a) \geq \max(a)$ could be used as *f*. However, in that case regular, row- and column-regular blocks are no longer compatible with complete blocks and binary blockmodeling is no longer a special case of valued blockmodeling.

The block inconsistencies presented in Table 2 are very similar to those suggested by Batagelj and Ferligoj (2000). There are three main differences:
1. Batagelj and Ferligoj do not use the parameter *m*. In their approach, the parameter *m* is replaced by the maximum of the block analyzed. This makes their criterion function compatible with structural and regular equivalence, defined in Section 2. However, this also makes their criterion function strongly dependent on the block maximums.
2. In the approach by Batagelj and Ferligoj, *f* is fixed at *maximum* (for example in *f*-regular blocks).
3. Their block inconsistencies are normalized to the values on the interval from 0 to 1.

**The parameter *m***

The main problem is to determine an appropriate value of the parameter *m*. The parameter *m* presents the minimal value that characterizes the tie between a unit and either a cluster (for *f*-regular, row-*f*-regular and column-*f*-regular blocks) or another unit (for complete, row-dominant, column-dominant, row-functional and column- functional blocks) in such a way that this tie satisfies the condition of the block.

---
[8] The exception is the regular block, where the weighted sum is corrected for the overlap of errors in rows and columns.

*Table 2: Block inconsistencies for binary and valued blockmodeling*

Block inconsistencies - $\delta(R(C_a, C_b), T)$

| Ideal block | Binary blockmodeling[9] | Valued blockmodeling | Position of the block |
|---|---|---|---|
| null | $s_t$ | $\sum_{i=1}^{n_r}\sum_{j=1}^{n_c} b_{ij}$ | *off-diagonal* |
|  | $s_t + min(0, n_r - 2s_d)$ | $\sum_{i=1}^{n_r}\sum_{j=1}^{n_c} b_{ij} + \min\left(0, \sum(m-diag(B))^+ - \sum diag(B)\right)$ | *Diagonal* |
| complete | $n_r n_c - s_t$ | $\sum_{i=1}^{n_r}\sum_{j=1}^{n_c} (m - b_{ij})^+$ | *off-diagonal* |
|  | $n_r n_c - s_t + min(-n_r + 2s_d, 0)$ | $\sum_{i=1}^{n_r}\sum_{j=1}^{n_c} (m - b_{ij})^+ + \min\left(-\sum(m-diag(B))^+ + \sum diag(B), 0\right)$ | *Diagonal* |
| row-dominant | $(n_c - m_r) n_r$ | $\min\left((m-B)^+ \cdot 1\right) n_r$ | *off-diagonal* |
|  | $[n_c - m_r - (1 - s_d)^+] n_r$ | $\min\left((m-B)^+ \cdot 1 + \left(\sum diag(B) - (diag(m-B)^+)^-\right)\right) n_r$ | *Diagonal* |
| column-dominant | $(n_r - m_c) n_c$ | $\min\left(1'\cdot(m-B)^+\right) n_c$ | *off-diagonal* |
|  | $[n_r - m_c - (1 - s_d)^+] n_c$ | $\min\left(1'\cdot(m-B)^+ + \left(\sum diag(B) - (diag(m-B)^+)^-\right)\right) n_c$ | *Diagonal* |
| row-*f*-regular | $(n_r - p_r) n_c$ | $\sum_{i=1}^{n_r} (m - f(B_{[i,]}))^+ n_c$ |  |
| column-*f*-regular | $(n_c - p_c) n_r$ | $\sum_{j=1}^{n_c} (m - f(B_{[,j]}))^+ n_r$ |  |
| *f*-regular | $(n_c - p_c) n_r + (n_r - p_r) p_c$ | $\sum_{i=1}^{n_r}\sum_{j=1}^{n_c} \max\left((m - f(B_{[i,]}))^+, (m - f(B_{[,j]}))^+\right)$ |  |
| row-functional | $s_t - p_r + (n_r - p_r) n_c$ | $\sum_{i=1}^{n_r}\left((m - \max(B_{[i,]}))^+ n_c + \sum_{j=1, j \neq \arg\max b_{ij}}^{n_r} b_{ij}\right)$ |  |
| column-functional | $s_t - p_c + (n_c - p_c) n_r$ | $\sum_{j=1}^{n_c}\left((m - \max(B_{[,j]}))^+ n_r + \sum_{i=1, i \neq \arg\max b_{ij}}^{n_r} b_{ij}\right)$ |  |

Legend:

$s_t$ - total block sum = number of 1s in a block; $s_d$ - diagonal block sum = number of 1s on a diagonal; $n_r$ - number of rows in a block = card $C_i$; $n_c$ - number of columns in a block = card $C_j$; $p_r$ - number of non-null rows in a block; $p_c$ - number of non-null columns in a block; $m_r$ - maximal row-sum, $m_c$ - maximal column-sum; B - the matrix of the block $R(C_i, C_j)$; $B_{[i,]}$ - the *i*-th row of the matrix B; $B_{[,j]}$ - the *j*-th column of the matrix B; $b_{ij}$ - an element of matrix B defined by i-th row and j-th column; diag - extract the diagonal elements of the matrix

$$(x)^+ = \begin{cases} x, & x > 0 \\ 0, & \text{otherwise} \end{cases} \qquad (x)^- = \begin{cases} x, & x < 0 \\ 0, & \text{otherwise} \end{cases}$$

---

[9] Slightly adapted from Doreian et al. 2005: 224.

For example, for a citation network and *m* equal to 5, an author is (strongly) linked to another author, if (s)he has cited the author at least 5 times.

The best way for determining the parameter *m* is prior knowledge, which can tell us how strong a tie should be to be considered strong or relevant. If such prior knowledge does not exist, the following guidelines can be helpful.

A partition obtained with another approach can be used. Given a specified blockmodel, an *m* is sought which approximately matches the partition. Such *m* is not the *m* that minimizes the total inconsistency, since this would always be achieved by setting *m* (close) to 0. For example, for complete blocks, the means of the complete blocks or the distribution of values in these blocks could be examined. For regular blocks, the mean or distribution of values of the function *f* over rows and columns in the regular blocks could be examined. This procedure should give an (interval) estimate of possible *m* values, which should then be tested.

The parameter *m* can also be determined using the distribution of appropriate values. The nature of these values depends on the ideal blocks in the desired blockmodel:
- for models without *f*-regular, row- and column-*f*-regular blocks (e.g. complete, dominant and null blocks), the distribution of cell values must be examined.
- for models with *f*-regular, row- and column-*f*-regular blocks, the distribution of row or column function *f* values (the values of function *f* computed over rows or columns) must be examined.

If a distribution (where the value 0 is excluded) is bimodal, the parameter *m* should be chosen somewhere in between the both modes. If the distribution is unimodal, the parameter *m* could be set around the mode. If the distribution of row or column function *f* values is considered and the function *f* is influenced by the number of units over which it is computed (such as *sum*), the expected number of *f*-regular or row- or column-*f*-regular blocks for at least some groups must be taken into account.

An appropriate *m* should also be somewhere between the threshold (slicing parameter) used for binarizing (slicing) the network and double that value. Both the slicing parameter and the parameter *m* distinguish between relevant and irrelevant ties. However, there is also an important difference. The slicing parameter makes the distinction between relevant and irrelevant ties in such a way that a tie is relevant if it is higher or equal to the slicing parameter, and irrelevant otherwise. On the other hand, a tie is considered relevant if it is closer to *m* than to 0. Therefore, *m* equal to double the slicing parameter would classify the same values as relevant and irrelevant. If *f*-regular, row- and column-*f*-regular blocks are used and the function *f* has a property $f(a) > \max(a)$, *m* can be even higher.

## 5 Homogeneity blockmodeling

Another approach to generalized blockmodeling of valued networks is to search for homogeneity within blocks. It searches for the partition where the sum of some measure of within block variability over all blocks is minimal. The idea was presented by Borgatti and Everett (1992b). The measure of variability measures the inconsistency of an empirical block with the ideal block. Based on this definition of block inconsistency, ideal blocks for homogeneity blockmodeling can be defined and incorporated into the framework presented previously (in the criterion function presented in Section 3).

Two measures of variability can be defined: the "sum of square deviations from the mean" (sum of squares) and the "sum of the absolute deviations from the median" (absolute deviations). The measures of variability in a block can be defined in several ways. The measure of variability can either be computed over all cell values in the block (for structural equivalence) or over values of the function *f* over rows, columns or both separately (for row-*f*-regular, column-*f*-regular and *f*-regular blocks[10]). If the measure is computed over the values of the function *f* over rows or columns, the result is then multiplied by the number of elements in each row or column respectively.

The descriptions of ideal blocks for homogeneity blockmodeling are presented in the last (fourth) column in Table 1. Here it can be seen that the null block (in homogeneity blockmodeling approach) is only a special case of the complete block. As always in generalized blockmodeling, the complete block is a special case of row-(*f*-)regular, column-(*f*-)regular and (*f*-)regular blocks[11]. In homogeneity blockmodeling, the null block is therefore also a special case of these blocks.

Based on these descriptions and a selected measure of variability, block inconsistencies can be defined. The block inconsistencies for both types of homogeneity blockmodeling are presented in the Table 3.

*Table 3: Block inconsistencies for homogeneity blockmodeling*

| Ideal block | Block inconsistencies - $\delta(R(C_a,C_b),T)$ | | Position of the block |
|---|---|---|---|
| | sum of squares | absolute deviations | |
| null | $\sum_{i,j} b_{ij}^2$ | $\sum_{i,j} |b_{ij}|$ | off-diagonal |
| | $\sum_{i \neq j} b_{ij}^2 + ss(diag(B))$ | $\sum_{i \neq j} |b_{ij}| + ad(diag(B))$ | diagonal |
| complete | $ss_{i,j}(b_{ij})$ | $ad_{i,j}(b_{ij})$ | off-diagonal |
| | $ss_{i \neq j}(b_{ij}) + ss(diag(B))$ | $ad_{i \neq j}(b_{ij}) + ad(diag(B))$ | diagonal |
| row-*f*-regular | $ss_i(f(B_{[i,]}))n_c$ | $ad_i(f(B_{[i,]}))n_c$ | |
| column-*f*-regular | $ss_j(f(B_{[,j]}))n_r$ | $ad_j(f(B_{[,j]}))n_r$ | |
| *f*-regular | $\max\left(ss_i(f(B_{[i,]}))n_c, ss_j(f(B_{[,j]}))n_r\right)$ | $\max\left(ad_i(f(B_{[i,]}))n_c, ad_j(f(B_{[,j]}))n_r\right)$ | |

Legend:
B - matrix of block $R(C_i, C_j)$; $B_{[i,]}$ - the *i*-th row of the matrix B; $B_{[,j]}$ - the *j*-th column of the matrix B; $b_{ij}$ - an element of matrix B defined by *i*-th row and *j*-th column; $n_r$ - number of rows in a block = card $C_i$; $n_c$ - number of columns in a block = card $C_j$; *diag*(B) - a vector of the diagonal elements of the matrix B; *Me(x)* – median; $\bar{x}$ - arithmetic mean; *ss(x)* - sum of square deviations from the mean: $ss(x) = \sum_i (x_i - \bar{x})^2$ ;

*ad(x)* - sum of absolute deviations from the median: $ad(x) = \sum_i |x_i - Me(x)|$

---
[10] For discussion if any of these functions result in regular equivalence see Section 3.
[11] The complete block is also a special case of row- and column-dominant blocks. However these are for now not discussed within homogeneity approach.

Block inconsistencies for most of the ideal block presented in Table 3 follow quite naturally from the descriptions of ideal blocks in the last column of Table 1. The only exception is the block inconsistency for *f*-regular block.

Several possibilities were considered for the block inconsistency with the *f*-regular block (the last row in Table 3). Namely, sum and mean have been considered instead of maximum. Although the use of sum might seam the most logical, this might make the block inconsistency for *f*-regular block too large. Similar approach was taken in valued blockmodeling, however there a special care was taken in order to make sure that each cell can contribute only once to the *f*-regular block inconsistency. This approach is not possible here.

Maximum was eventually chosen, since it preserves the inequalities of block inconsistencies that holds in both binary and valued blockmodeling:

$$\delta(R(C_a,C_b),reg) \leq \delta(R(C_a,C_b),cre) \leq \delta(R(C_a,C_b),com) \text{ and}$$

$$\delta(R(C_a,C_b),reg) \leq \delta(R(C_a,C_b),rre) \leq \delta(R(C_a,C_b),com).$$

Row(column)-*mean*-regular (*f* is set to *mean*) and complete blocks are compatible. The block inconsistencies for "sum of squares complete block" and "sum of squares row(column)-*mean*-regular block" match if and only if the rows(columns) are homogeneous, that is if each row(column) has zero variance. Therefore, using *mean* for *f* is suggested always, when *f*-regular (, row- and column-*f*-regular) and complete blocks are used in the same blockmodel. Functions other than *mean* can be used for *f*; however, then the inconsistencies of *f*-regular (, row- and column-*f*-regular) and complete blocks are not compatible. In this case, such ideal blocks can not be used in the same blockmodel. The block inconsistencies can be adjusted for pre-specified blockmodeling by substituting mean ($\bar{x}$) or median (*Me(x)*) (as a value from which deviations are computed) by the pre-specified value. The block inconsistency for null block can be seen as an example how the inconsistency for complete block can be adjusted for the pre-specifies value of 0.

The main advantage of homogeneity blockmodeling over binary and valued blockmodeling is that it does not require any additional parameters (such as parameter *m* in valued blockmodeling). Therefore, solutions gained using homogeneity blockmodeling could be used as an initial solution for binary or valued blockmodeling. In addition, no information is lost using homogeneity blockmodeling and negative values of the ties can be used directly.

A similar approach for structural equivalence was already suggested by Borgatti and Everett (1992b). They used average variance within matrix blocks as a measure of fit. The disadvantage of this measure compared with the sum of squares and absolute deviations measures proposed in this paper is that the size of the block has a large effect on the contribution of an individual cell (in the matrix) to the total block inconsistency.

Under the generalized blockmodeling approach presented by Doreian et al. (2005), block inconsistencies can also be "normalized" by dividing them by the number of cells in a block. In this case, the sum of squares approach (blockmodeling) changes into the variance approach.

# 6 Example: Notes borrowing between social-informatics students (line measurement)

The data in this example come from a survey conducted in May 1993 on 13 social-informatics students (Hlebec, 1996). The network was constructed from answers[12] to the question, "How often did you borrow notes from this person?" for each of the fellow students. The results are presented in Figure 1.

*Figure 1: Valued network of notes borrowing between social-informatics students*

**Original network**

|    | 1 | 2 | 3  | 4  | 5 | 6 | 7  | 8  | 9  | 10 | 11 | 12 | 13 |
|----|---|---|----|----|---|---|----|----|----|----|----|----|----|
| 1  |   |   |    | 15 |   |   |    | 1  | 8  |    |    | 3  |    |
| 2  |   |   | 2  | 3  |   |   | 5  | 5  | 10 | 10 | 1  | 3  |    |
| 3  |   |   |    | 19 |   |   |    | 3  | 1  |    |    |    |    |
| 4  | 2 |   | 6  |    | 1 |   |    | 1  | 19 |    | 1  |    |    |
| 5  |   |   |    | 16 |   | 5 |    | 7  | 16 |    | 5  |    | 3  |
| 6  |   |   | 1  |    | 4 |   |    | 7  | 3  |    | 7  | 3  | 1  |
| 7  |   |   | 6  | 14 |   |   |    | 14 | 6  |    |    |    |    |
| 8  |   |   |    | 5  |   |   |    |    | 6  |    |    |    |    |
| 9  |   |   |    | 19 |   |   |    | 1  |    |    |    |    |    |
| 10 |   |   | 16 | 2  | 16|   | 1  | 16 |    |    | 1  | 2  |    |
| 11 |   |   | 2  | 8  | 2 | 2 |    | 5  | 14 |    |    | 2  |    |
| 12 | 2 | 2 | 8  | 2  | 2 | 2 | 2  | 2  | 6  | 2  | 11 |    |    |
| 13 |   |   |    | 1  | 8 |   |    | 8  | 3  |    |    |    |    |

The aim of the analysis is to discover groups of students that play similar roles. All types of generalized blockmodeling discussed in the paper are considered and compared.

It seems unlikely that a student would borrow notes from each student in a given group. Therefore, the model of (*f*-)regular equivalence is used. Several functions are used as *f*. These functions are *maximum* and *sum* for valued blockmodeling and *maximum* and *mean* for homogeneity blockmodeling. The use of *maximum* is based on the ideas of Batagelj and Ferligoj (2000) and the REGGE algorithm (White, 1985a). There *maximum* is presented as a suitable function for generalization of regular equivalence to valued networks. The function *sum* is used in valued blockmodeling since it is assumed that students of a given group want to borrow a certain amount of notes from students of another (or the same) group. The problem with this assumption is that the values do not represent the amount of notes borrowed. The use of function *mean* in homogeneity blockmodeling is equivalent to the use of function *sum* in valued blockmodeling.

It should be noted that the three main types of generalized blockmodeling use slightly different definitions of (*f*-)regular blocks, even if the same *f* is used in *f*-regular blocks. For the use of binary blockmodeling, a valued network must first be converted into a binary one. This is done by recoding all values lower than the slicing parameter into zeros and the rest into ones. Then the definition of a regular block is that the maximum of each row and each column must be at least the slicing parameter. If the slicing parameter is replaced with *m*, the result is the definition of *max*-regular blocks for valued blockmodeling. The definition of *max*-regular block in homogeneity blockmodeling is slightly different. Here the maximums of all rows and all columns separately must be equal. The definitions for other functions as *f* can be formulated based on Table 1.

---

[12] The respondents indicated the frequency of borrowing by choosing (on a computer) a line of length 1 to 20, where 1 meant no borrowing. 1 was deducted from all answers, so that 0 now means no borrowing.

By inspecting several partitions produced by different types of generalized blockmodeling, allowed ideal blocks and with different numbers of clusters, it can be seen that a partition into three clusters is the most appropriate. Therefore, only partitions into 3 clusters are presented. The remainder of this section follows the procedure suggested in Section 7. There the rational for such procedure is also explained.

**Homogeneity blockmodeling**

First homogeneity blockmodeling was applied to the network.

Optimal partition for sum of squares and absolute deviations blockmodeling [13] with *mean*-regular blocks is presented in Figure 2.

*Figure 2: Optimal partitions for sum of squares and absolute deviations homogeneity blockmodeling with mean-regular blocks*

**Sum of squares and Absolute deviations**
*mean*-regular

|    | 1 | 5 | 7 | 10 | 11 | 2 | 3 | 6 | 12 | 13 | 4 | 8 | 9 |
|----|---|---|---|----|----|---|---|---|----|----|---|---|---|
| 1  |   |   |   |    |    |   |   | 3 |    |    | 15| 1 | 8 |
| 5  |   |   |   |    | 5  |   | 5 |   |    | 3  | 16| 7 | 16|
| 7  |   |   |   |    |    | 6 |   |   |    |    | 14| 14| 6 |
| 10 |   |   |   |    | 1  | 16| 2 | 1 | 2  |    | 16| 16|   |
| 11 |   | 2 |   |    |    |   | 2 | 2 | 2  |    | 8 | 5 | 14|
| 2  |   | 5 | 10| 1  |    |   | 2 |   | 3  |    | 3 | 5 | 10|
| 3  |   |   |   |    |    |   |   |   |    |    | 19| 3 | 1 |
| 6  |   | 4 |   |    | 7  |   | 1 |   | 3  | 1  |   | 7 | 3 |
| 12 | 2 | 2 | 2 | 2  | 11 | 2 | 8 | 2 |    |    | 2 | 2 | 6 |
| 13 |   | 8 |   |    |    |   |   |   |    |    | 1 | 8 | 3 |
| 4  | 2 | 1 |   |    | 1  |   | 6 |   |    |    |   | 1 | 19|
| 8  |   |   |   |    |    |   |   |   |    |    | 5 |   | 6 |
| 9  |   |   |   |    |    |   |   |   |    |    | 19| 1 |   |

Based on this partition, the appropriate *m* for valued blockmodeling with null and *sum*-regular blocks would be around 5 or 10. This assumption is based on the following two matrices that represent the mean row and columns sums in each block of that blockmodel.
Mean row sums:
```
        1    2    3
1     2.0  8.8 31.2
2    10.8  5.5 14.6
3     1.3  2.0 25.5
```

Mean column sums:
```
        1    2    3
1     2.0  8.8 52.0
2    10.8  5.5 24.3
3     0.8  1.2 25.5
```

---

[13] In this case both types of homogeneity blockmodeling produce the same partition.

However instead of the function *mean* over rows and columns, *maximum* could be also used, or to put it differently, instead of searching for *mean*-regular blocks, *max*-regular blocks could be searched for. The optimal partitions for *max*-regular blocks are presented in Figure 3.

*Figure 3: Optimal partitions for homogeneity blockmodeling with max-regular blocks*

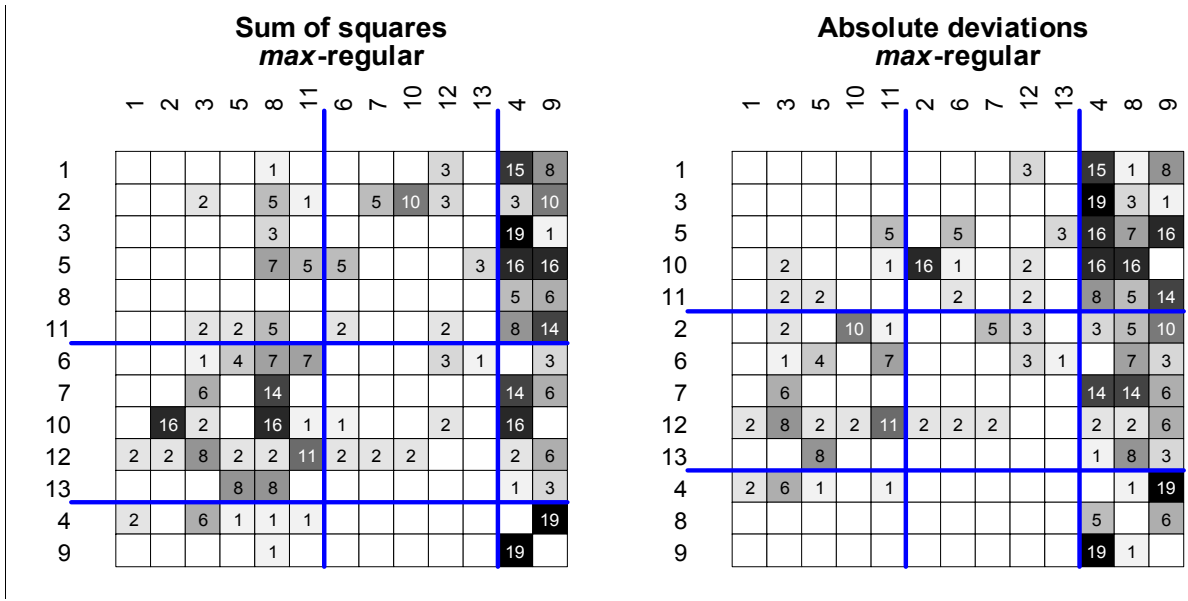

Here the sum of squares and absolute deviations partitions do not match. Subjective judgment is needed to determine which partition is better. Absolute deviations partition seems better. It induces blocks that seem "cleaner". The upper and lower left blocks can be now more easily interpreted as null blocs[14] and the unit 8 fits quite nicely in the third group. Based on this partition, the appropriate *m* for valued blockmodeling with null and *max*-regular blocks would be around 5. This assumption is based on the following two matrices that represent the mean row and columns maximums in each block of that blockmodel.

Mean row maximums:
```
      1    2    3
1   1.8  5.2 16.0
2   8.4  2.0  9.0
3   2.0  0.0 14.7
```

Mean column maximums:
```
      1    2    3
1   1.8  5.4 17.0
2   7.8  2.6 12.7
3   2.0  0.0 13.0
```

**Valued blockmodeling**

The best[15] results for valued blockmodeling using null and *sum*-regular blocks were obtained with one of the suggested *m* values - 10. It produced the partition in Figure 4 and the following model (since it was already stated that *sum*-regular blocks were used, only "reg" is used to indicate *sum*-regular blocks):
```
      1       2      3
1  "null"  "null"  "reg"
2  "null"  "reg"   "reg"
3  "null"  "null"  "reg"
```

---

[14] The procedure identifies these blocks as *f*-regular blocks. As noted in Section 5, null blocks are only a special case of *f*-regular blocks, where the value of function *f* for all rows and all columns is exactly 0. This rarely happens. However, when we are interpreting the result of the blockmodel, we can interpret *f*-regular blocks that are close to null blocks as null blocks.

[15] Again based on subjective judgment.

For valued blockmodeling with null and *max*-regular blocks, *m* equal to 5 performed better, again as suggested by *max*-regular absolute deviations partition. This partition is identical to the one obtained using *sum*-regular equivalence and therefore presented in Figure 4. The model is also the same as above (only that "reg" now indicates *max*-regular blocks).

*Figure 4: Optimal partition for valued blockmodeling obtained using null and max-regular blocks with m equal to 5 or using null and sum-regular blocks with m equal to 10*

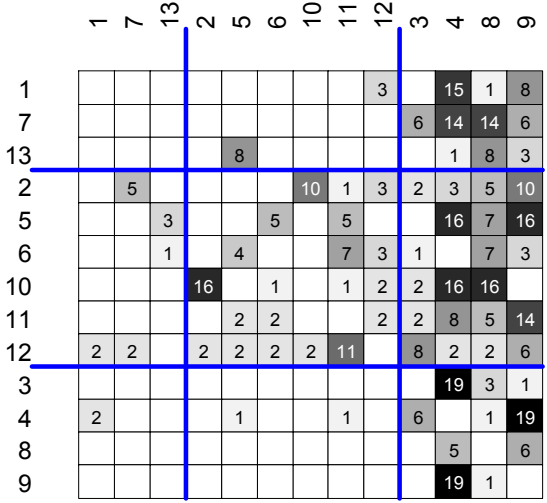

**Binary blockmodeling**

In this case binary blockmodeling does not produce as good results as the previous approaches. It was explored on the matrix on Figure 1 sliced at 1, 2, 3, 5, and 10 (values of ties equal or grater to this values were recoded into ones). These values were suggested by the barplot on Figure 5.

*Figure 5: Histogram of cell values for the matrix on Figure 1*

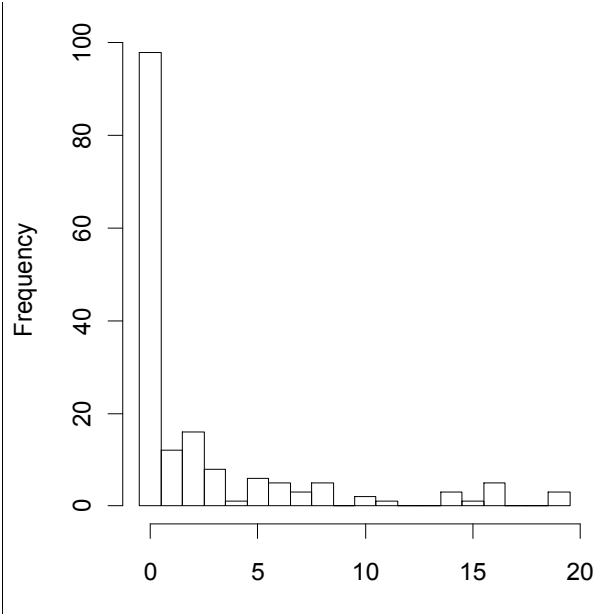

All of the 3 cluster solutions contain obvious misclassifications[16]. The best among them are presented in Figure 6. The first solution is based on the network sliced at 1. It is presented on the first matrix in Figure 6 as a binary network and on the second matrix as a valued network. The biggest problem with this solution is that the unit 4 is not in the first group. This might not be so evident if we would be looking at a binary network, but with additional data that the valued network provides, it is obvious. The same setting (matrix on Figure 1 sliced at 1) produced another optimal solution, which is presented on the third matrix in Figure 6. These two solutions (based on network sliced at 1) have the following images:

```
Solution 1                        Solution 2
       1      2      3                   1      2      3
1  "null"  "reg"  "reg"          1  "null"  "reg"  "reg"
2   "reg"  "reg"  "reg"          2   "reg"  "reg"  "reg"
3  "null" "null"  "reg"          3  "null" "null"  "reg"
```

*Figure 6: Optimal partitions for binary blockmodeling with regular equivalence with different slicing parameters*

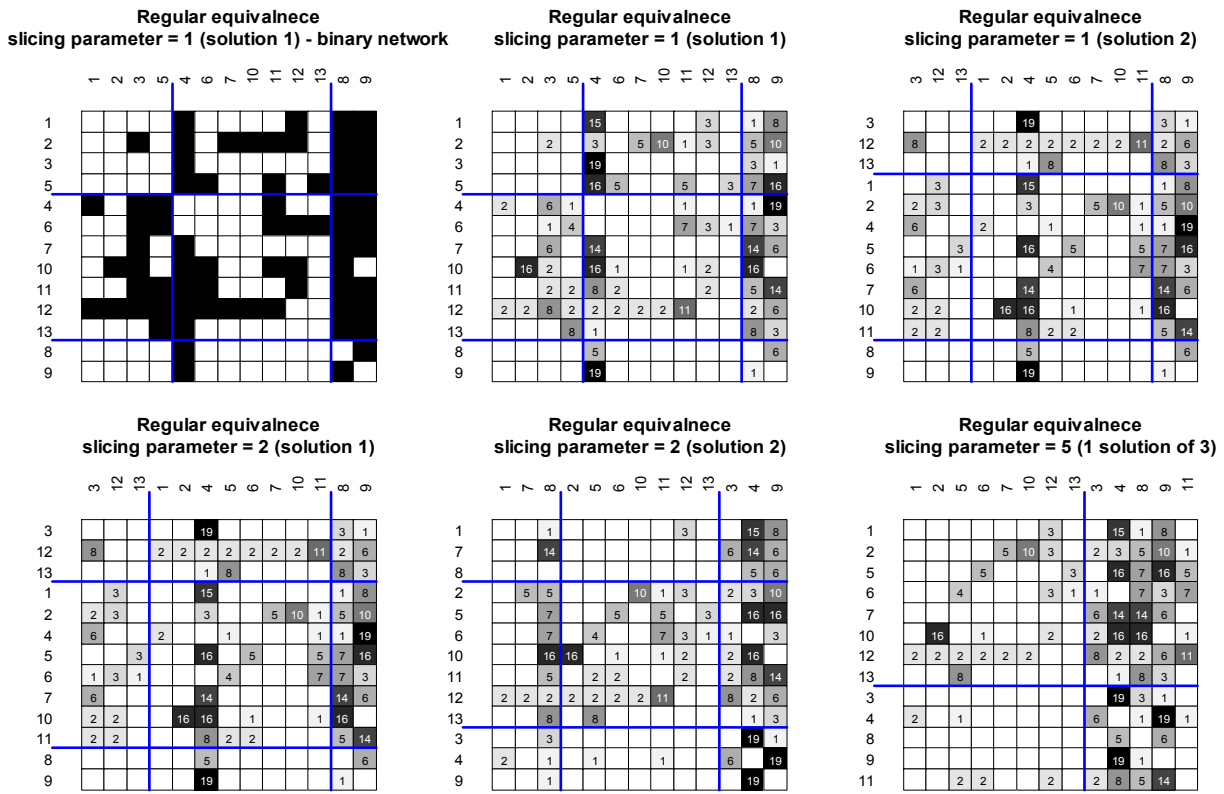

Partitions presented on the fourth and fifth matrices in Figure 5 were obtained on the network sliced at 2. The partition of the second solution is the same as the one of the second solution obtained on network sliced at 1. The solutions obtained on network sliced at 2 are reasonably good, although it is obvious that at least unit 8 in the first solution and unit 4 in the second solution are misclassified. The accompanying images are:

---

[16] The 2 cluster solution has other deficiencies.

Solution 1
```
    1       2       3
1 "null"  "reg"   "reg"
2 "reg"   "reg"   "reg"
3 "null"  "null"  "null"
```

Solution 2
```
    1       2       3
1 "null"  "null"  "reg"
2 "reg"   "reg"   "reg"
3 "null"  "null"  "reg"
```

The partition on the sixth matrix in Figure 6 is based on a network sliced at 3 and has the image:
```
    1       2       3
1 "null"  "null"  "reg"
2 "null"  "reg"   "reg"
3 "null"  "null"  "null"
```

As in the previous partitions, at least one unit (unit 4) is misclassified. A search for 3 cluster solution on network sliced at 5 did not produce satisfactory results. It produced 100 partitions with minimum inconsistency.

**Interpretation**

It is hard to select the most appropriate partition out of the ones presented, since they were obtained using different criteria functions and different definitions of ideal blocks. A researcher can select a partition based on several criteria, some of which are:
  o   the suitability of the definition of ideal blocks,
  o   the possibility of logical interpretation of the partition and obtained blockmodel, and
  o   the "aesthetic" characteristics of the partition.

The partitions that seem the most suitable are both partition based on absolute deviations blockmodeling (with *max*- and *mean*-regular blocks) and the partition based on valued blockmodeling (which was obtained with a model allowing null blocks and either *max*-regular blocks with $m = 5$ or *sum*-regular blocks with $m = 10$).

In all these partitions, a group exists from which everybody borrows notes and whose members do not borrow notes from anybody outside their group. This group consists of units 4, 8, 9 and sometimes unit 3. This is probably the group that makes good notes on a regular basis (presumably good students).

The interpretation of the other two groups differs depending on which of the solutions we interpret. Based on the partitions obtained using absolute deviations blockmodeling, the other two groups exhibit similar behavior. They borrow notes from both remaining groups (besides themselves) and lend notes to each other; however, they only seldom borrow (lend) notes within their group. The main difference between these two groups is that one of them relies more heavily on the first group (good students). This might be two groups of students that do not have much contact with the members of their group.

The interpretation of the valued blockmodeling solution seems more logical. Again there are two groups in addition to the good students group. Both of them borrow notes from the good students group, while only one of them borrows within its group. They rarely borrow from each other. One of them could be a group of students that do not have much contact with fellow students and the other the "average" students.

# 7 Suggested procedure for generalized blockmodeling of valued networks

Two main approaches or types of generalized blockmodeling of valued networks have been suggested. In this section some suggestions are made about the selection of a type of generalized blockmodeling for a particular generalized blockmodeling problem. These suggestions were already followed in the example in the previous section.

The main suggestion is that if it is possible, both valued and homogeneity blockmodeling should be considered. Of course, this is a very general suggestion. More complete guidelines have to take into account:
- o the nature of data, the prior knowledge that exists about the studied network or specified blockmodel,
- o the type of equivalence or ideal blocks sought, and
- o does a pre-specified model exist and if so, how is it specified.

Most of the suggestions presented in this section follow quite naturally from the advantages and disadvantages described in the following section.

The selection of the appropriate approach is usually quite straightforward in the case of sufficient prior knowledge and optional pre-specified models. Therefore this situation is not covered in this section. Also the selection of the approach is meaningless when only one approach is appropriate for either the data or selected ideal blocks. For example, at least for now, valued blockmodeling can not handle negative values of the ties[17]. On the other hand, the inconsistencies for row- and column-dominant and row- and column-functional blocks are currently not defined for homogeneity blockmodeling.

What is described here is a suggested procedure for generalized blockmodeling of valued networks when:
- o both approaches are feasible (meaning especially only nonnegative values of the ties and ideal blocks supported by homogeneity blockmodeling) and
- o no, or at least insufficient, prior knowledge is available.

In this case, homogeneity approach should be applied first with appropriately selected ideal blocks. Both types of homogeneity blockmodeling can be tested. Then valued blockmodeling can be used to check if any improvement is possible, especially if more complex models are desired than supported by homogeneity blockmodeling.

For valued blockmodeling, the parameter $m$ is required. Since a partition is already available, this information can be used to select an appropriate $m$. For example, for complete blocks, the means of the complete blocks or the distribution of values in these blocks should be examined. For regular blocks, the mean or distribution of values of the function $f$ over rows and columns in the regular blocks should be examined. This procedure should give an (interval) estimate of possible $m$ values, which should then be tested using valued blockmodeling.

# 8 Advantages and disadvantages of proposed approaches

First, the advantages and disadvantages that are common to both proposed approaches to generalized blockmodeling of valued networks (valued and homogeneity blockmodeling) are listed and described. Then, the advantages and disadvantages specific to only one of the types of generalized blockmodeling are discussed.

---

[17] The valued blockmodeling could be easily adapted to allow negative values of ties. Even without adaptations it is sometimes possible to convert a network so that all values are nonnegative and then apply valued blockmodeling.

**Advantages and disadvantages of generalized blockmodeling of valued networks**

The main advantage of the blockmodeling of valued networks compared to binary blockmodeling is that more information is used[18]. This is actually the source of the other advantage.

The second advantage is that there are fewer partitions having identical values of the criterion function. It could be said that approaches for generalized blockmodeling of valued networks measure the inconsistencies more precisely. Only one optimal partition is usually found for generalized blockmodeling of valued networks, especially for homogeneity blockmodeling. This can be also seen in the example in Section 6. More then one optimal partition occurred on several occasions when using binary blockmodeling and never while using valued or homogeneity blockmodeling.

A requirement of the proposed approaches is that the network should be measured on at least interval scale. This may (or may not) be a disadvantage.

**Advantages and disadvantages of valued blockmodeling**

The need for the parameter $m$ is a disadvantage of valued blockmodeling, but only with respect to homogeneity blockmodeling. It is not a disadvantage with respect to binary blockmodeling, as a slicing parameter is used in binary blockmodeling for binarization of the valued networks. The valued blockmodeling is less sensitive to the selection of the parameter $m$ than binary blockmodeling to the selection of the slicing parameter. In valued blockmodeling, in e.g. complete blocks, values of ties just under $m$ have only a small inconsistency, while in binary blockmodeling all values of ties under the slicing parameter have equal inconsistencies. This is even more true in the case of *sum*-regular blocks, since values in a row or column in a block that are too small to be significant by themselves, can sum up to a significant relation between an individual and a group.

Valued and binary blockmodeling have one advantage compared to homogeneity blockmodeling. This advantage is the richness of possible ideal blocks that they allow. Homogeneity blockmodeling, at least for now, has a very limited set of allowed ideal blocks[19]. However, this might change in the future, since other ideal blocks could probably be adapted for homogeneity blockmodeling as well. Even within this limited set of ideal blocks, all are not necessary compatible. Actually, in homogeneity blockmodeling, *f*-regular, row- and column-*f*-regular blocks are only compatible with complete blocks, if the *f* (the function used in *f*-regular, row-, and column-*f*-regular blocks) is *mean*.

Another disadvantage of valued blockmodeling is the way block inconsistencies are computed. The null blocks have a handicap in the presence of large values. While for complete blocks each cell can contribute to a block inconsistency at maximum $m$, in null blocks this contribution is not limited. The problem with this is that a large enough value can cause an otherwise null block to be declared complete. This problem could be overcome by censoring[20] the network at some value and thus limiting the contribution of each cell to block inconsistency for null block to this value.

---

[18] Using these approaches, no information or only a little is lost (See Section 4 for discussion where information is lost or discarded.).
[19] See Table 5.1.
[20] By censoring is meant recoding values over some threshold to this value. The threshold could be $m$ or higher. By selecting threshold $m$, the maximum contribution of a cell to block inconsistency would be $m$ for both null

Although valued blockmodeling uses substantially more information than binary blockmodeling, it usually still discards some information. For example, when computing inconsistencies for complete blocks, it discards information about cell values above *m*. Homogeneity blockmodeling, on the other hand, does not discard any information. This also means that valued blockmodeling (or its criterion function) is not fully compatible with the definitions of structural and *f*-regular equivalences defined in Section 2.

**Advantages and disadvantages of homogeneity blockmodeling**

The main advantage of homogeneity blockmodeling is that no additional parameters need to be set in advance and its main disadvantage is that it can consider fewer possible ideal blocks than both binary and valued blockmodeling. In addition, only homogeneity blockmodeling uses all available information. As noted above, homogeneity blockmodeling (or its criterion function) is also fully compatible with the definitions of structural and *f*-regular equivalences defined in Section 2.

# 9 Conclusions

In the paper, two new approaches to generalized blockmodeling of valued networks have been presented. Several types of generalized blockmodeling are distinguished: binary blockmodeling, valued blockmodeling and homogeneity blockmodeling. The term binary blockmodeling is used for generalized blockmodeling of binary networks, presented by Doreian et al. (2005).

The two new approaches use more information (values of ties, not only on existence of a tie) about the network than binary blockmodeling. As a result, they produce fewer partitions with identical values of the criterion function. The requirement of the proposed approaches is that the network ties should be measured on at least interval scale.

Valued blockmodeling is a straightforward generalization of binary blockmodeling, with binary blockmodeling a special case of valued blockmodeling. Valued blockmodeling is less influenced by the initial parameters than binary blockmodeling (the parameter *m* vs. the slicing parameter) for valued data. Valued blockmodeling can also consider more ideal blocks than homogeneity blockmodeling and has fewer problems with compatibility of ideal blocks. These are the advantages of valued blockmodeling. There are also disadvantages. The parameter *m* must be set in advance while homogeneity blockmodeling requires no additional parameters. Valued blockmodeling (and binary blockmodeling) usually does (do) not use all available information on values of ties. In addition, values of ties considerably larger than the parameter *m* can severely penalize null blocks. This can distort a solution. However, this disadvantage can be overcome by censoring large values.

The second approach presented is homogeneity blockmodeling. The basic idea of homogeneity blockmodeling is that the inconsistency of an empirical block with its ideal block can be measured by within block variability of appropriate values[21]. It addresses one of the main problems of valued and binary blockmodeling when applied to valued networks. When using valued blockmodeling, the parameter *m* must be set. This parameter tells us how strong a tie must be to be treated as relevant. A similar parameter must also be set when using the binary blockmodeling, although it is sometimes implicitly set to the minimum positive

---

and complete blocks. This threshold should not be lower than *m*, since this would make complete blocks with inconsistency 0 impossible.

[21] What the appropriate values are is determined by the ideal block to which inconsistencies for a selected empirical block are computed. These values are always based on values of the ties in the selected empirical block.

value (all ties with values higher than 0 are treated as relevant). Homogeneity blockmodeling requires no such parameters. In addition, only homogeneity blockmodeling uses all available information. However, it lacks the richness of possible ideal blocks that the valued and binary blockmodeling possess.

The definition of regular equivalence for valued networks was also discussed. This is one of the areas that needs further attention. The other would be a comparison of these approaches with an approach for generalized blockmodeling of valued networks suggested by Batagelj and Ferligoj (2000), and other approaches to blockmodeling of valued networks. The comparison with two versions of REGE algorithm (White, 1985a; 1985b) is discussed in Žiberna (2005).

# 10 References:

Batagelj, V., 1997. Notes on Blockmodeling. Social Networks 19, 143-155.

Batagelj, V., Doreian, P., Ferligoj, A., 1992a. An optimizational approach to regular equivalence. Social Networks 14, 121–135.

Batagelj, V., Ferligoj, A., 2000. Clustering relational data. In: Gaul, W., Opitz, O., and Schader, M. (Eds.), Data Analysis, Springer-Verlag, New York, pp. 3 – 16.

Batagelj, V., Ferligoj, A., Doreian, P., 1992b. Direct and indirect methods for structural equivalence. Social Networks 14, 63–90.

Batagelj, V., Mrvar, A., Ferligoj, A., Doreian, P. (2004). Generalized Blockmodeling with Pajek. Metodološki zvezki, 1, 455-467.

Borgatti, S.P., Everett, M.G., 1992a. Notations of positions in social network analysis. In: Marsden, P.V. (Ed.), Sociological Methodology 22, Jossey-Bass, San Francisco, pp. 1–35.

Borgatti, S.P., Everett, M.G., 1992b. Regular blockmodels of multiway, multimode matrices. Social Networks 14, 91–120.

Breiger, R.L., Boorman, S., Arabie, P., 1975. An algorithm for clustering relational data with applications to social network analysis. Journal of Mathematical Psychology 12, 329–383.

Breiger, R.L., Mohr, J.W., 2004. Institutional Logics from the Aggregation of Organizational Networks: Operational Procedures for the Analysis of Counted Data. Computational & Mathematical Organization Theory 10, 17–43

Burt, R.S., 1976. Positions in networks. Social Forces 55, 93–122.

Doreian, P., 1988. Equivalence in a social network. Journal of Mathematical Sociology 13, 243–282.

Doreian, P., Batagelj, V., Ferligoj, A., 1994. Partitioning Networks on Generalized Concepts of Equivalence. Journal of Mathematical Sociology 19, 1–27

Doreian, P., Batagelj, V., Ferligoj, A., 2004., Generalized blockmodeling of two-mode network data. Social Networks 26, 29–53.

Doreian, P., Batagelj, V., Ferligoj, A., 2005. Generalized Blockmodeling. Cambridge University Press, New York.

Doreian, P., Mrvar, A., 1996. A partitioning approach to structural balance. Social Networks 18, 149–168.

Everett, M.G., Borgatti, S.P., 1994. Regular equivalence: General theory. Journal of Mathematical Sociology 191, 29–52.

Hlebec, V., 1996. Metodološke značilnosti anketnega zbiranja podatkov v analizi omrežji: Magistersko delo. FDV, Ljubljana.

Lorrain, F., White, H.C., 1971. Structural equivalence of individuals in social networks. Journal of Mathematical Sociology 1, 49–80.

White, D.R., 1985a. DOUG WHITE'S REGULAR EQUIVALENCE PROGRAM. http://eclectic.ss.uci.edu/~drwhite/REGGE/REGGE.FOR, accessed 12.5.2005.

White, D.R., 1985b. DOUG WHITE'S REGULAR DISTANCES PROGRAM. http://eclectic.ss.uci.edu/~drwhite/REGGE/REGDI.FOR, accessed 12.5.2005.

White, D.R., Reitz, K. P., 1983. Graph and semigroup homomorphisms on networks of relations. Social Networks 5, 193-234.

Žiberna, A., 2005. Direct and indirect methods for regular equivalence in valued networks. Paper presented at Applied Statistics conference, Ribno, Slovenia, 2005.

Žiberna, A. (2006). blockmodeling: An R package for generalized and classical blockmodeling of valued networks. http://www2.arnes.si/~aziber4/blockmodeling/, 15.3.2006.